\documentclass[useAMS,usenatbib,usegraphicx]{mn2e}

\usepackage{journals}

\title[Infrared identification of 4U1323-619 revisited]{Infrared identification of 4U1323-619 revisited}
\author[Zolotukhin, Revnivtsev, and Shakura]{Ivan Yu. Zolotukhin$^{1}$\thanks{E-mail:
iz@sai.msu.ru (IZ)}, Mikhail G.
Revnivtsev$^{2,3}$, and Nikolai I. Shakura$^{1}$\\
$^{1}$Sternberg Astronomical Institute, Moscow State University, Universitetskij pr., 13, 119992, Moscow, Russia\\
$^{2}$Excellence Cluster Universe, Technische Universit\"at M\"unchen, Boltzmannstr.2, 85748 Garching, Germany\\
$^{3}$Space Research Institute, Russian Academy of Sciences, Profsoyuznaya 84/32, 117997 Moscow, Russia
}
\begin{document}

\date{Accepted 2009 August 29. Received 2009 August 29; in original form 2009 August 11}

\pagerange{\pageref{firstpage}--\pageref{lastpage}} \pubyear{2009}

\maketitle

\label{firstpage}

\begin{abstract}
We re-examine the infrared counterpart of the dipping low-mass x-ray binary
4U1323-619. New X-ray data available from the {\it XMM} and {\it Chandra}
observatories combined with archival IR observations from the ESO 3.6m New
Technology Telescope allow us to define a new possible counterpart. We 
present here its photometric properties and compare them with a simple
analytical model of an accretion disc illuminated by the hot central
corona known to be present in the binary system.
\end{abstract}


\section{Introduction}

4U1323-619 is a dipping low-mass x-ray binary detected by {\it
Uhuru} and {\it Ariel V} \citep{forman78,warwick81}. It exhibits X-ray
bursts and irregular intensity dips repeating every 2.94~h. 
Dips discovered in EXOSAT data \citep{parmar89} lasting typically
$~30$~per~cent of the orbital cycle, and the lack of X-ray
eclipses are indicative that the binary plane is viewed at inclination 
$i = 60-80\degr$ \citep{frank87}. Such dipping behaviour is explained by
periodic obscurations of the central X-ray emitting region by a structure
located in the outer regions of the disc
\citep{white82}. High interstellar matter column density ($N_H\sim4 \times
10^{22}$~cm$^{-2}$) derived from X-ray spectral models
\citep{parmar89} suggests that this source possesses significant
extinction of the order of $A_V\sim18$~mag and, thus, cannot be observed at optical
wavelengths. \citet{smale95} attempted to find its IR counterpart
subject to much lower extinction and reported a candidate which
demonstrated some variability of the IR flux (though at a low
significance level). 

Similarly to other sources, the dipping activity in 4U1323-619
attracted attention of researchers as a laboratory for studying the
accretion disc corona \citep{balucinska99,boirin05,church05}. These
studies and other recent observational efforts produced new data
available for this source in archives of X-ray missions and the European
Southern Observatory (ESO). We have noticed a significant positional
discrepancy between the {\it Einstein} coordinates of 4U1323-619 and its
more recent {\it XMM} and {\it Chandra} observations. Having found the IR
archival data for 4U1323-619 by the means of the Virtual Observatory,
we decided to re-examine the IR candidate discovered by \citet{smale95}. 

\section{Observations and results}

\subsection{X-ray observations}

We have analysed {\it Chandra} observations performed on Sep 25, 2003
with a total exposure time about 40~ksec. The observations were aimed at high
energy resolution spectroscopy and the telescope was equipped with
gratings. Detectors were operating in the so-called Continuous-Clocking
 (CC) mode, when the information in one spatial direction is lost.

Therefore we have determined the source coordinates mainly in one direction
(with the nominal {\it Chandra} localisation accuracy $\sim 0.6$~arcsec at 
90~per~cent confidence). Orientation of 
the effective position ``strip'' depends 
on the orientation of the telescope during the observations. We represent
the {\it Chandra} localisation by a wide ellipse in Fig.~\ref{error_boxes}.
The centre of the ellipse is at RA(J2000)=201.65397~deg, Dec(J2000)=-62.135403~deg,  
its positional angle is 156.6~deg, the ellipse is $\sim 0.6$~arcsec wide 
in one direction and effectively infinitely long in another. In Fig.~\ref{error_boxes}, we have limited the size by arbitrarily adopting a length of 5~arcsec.

{\it XMM-Newton} observed the source position two times, on Aug 17, 2001
and Jan 29, 2003, the observations spanning 26 ksec and 51 ksec, respectively. Data of the {\it XMM-Newton} 
EPIC-MOS cameras were analysed using the standard tasks of the {\it Science 
Analysis Software (SAS) v9.0.0}.
Unfortunately, the source is sufficiently bright to have a non-negligible 
effect on the {\it XMM}/EPIC-MOS imaging capability because of pile-up. This depresses the brightness of central pixels, degrading the localisation accuracy, in spite of the possibility of cross-calibrating the absolute astrometry of {\it XMM} using optical sources within its field of view.
Therefore, we would like to adopt some conservative
uncertainty radius of the source position $\sim 3$~arcsec approximately 
corresponding to 90~per~cent confidence, which without pile-up might
have been significantly improved. 
The source position was determined to be RA(J2000)=201.654870~deg, 
Dec(J2000)=-62.135985~deg.

\subsection{IR observations}

\begin{figure*}
\includegraphics[width=\textwidth]{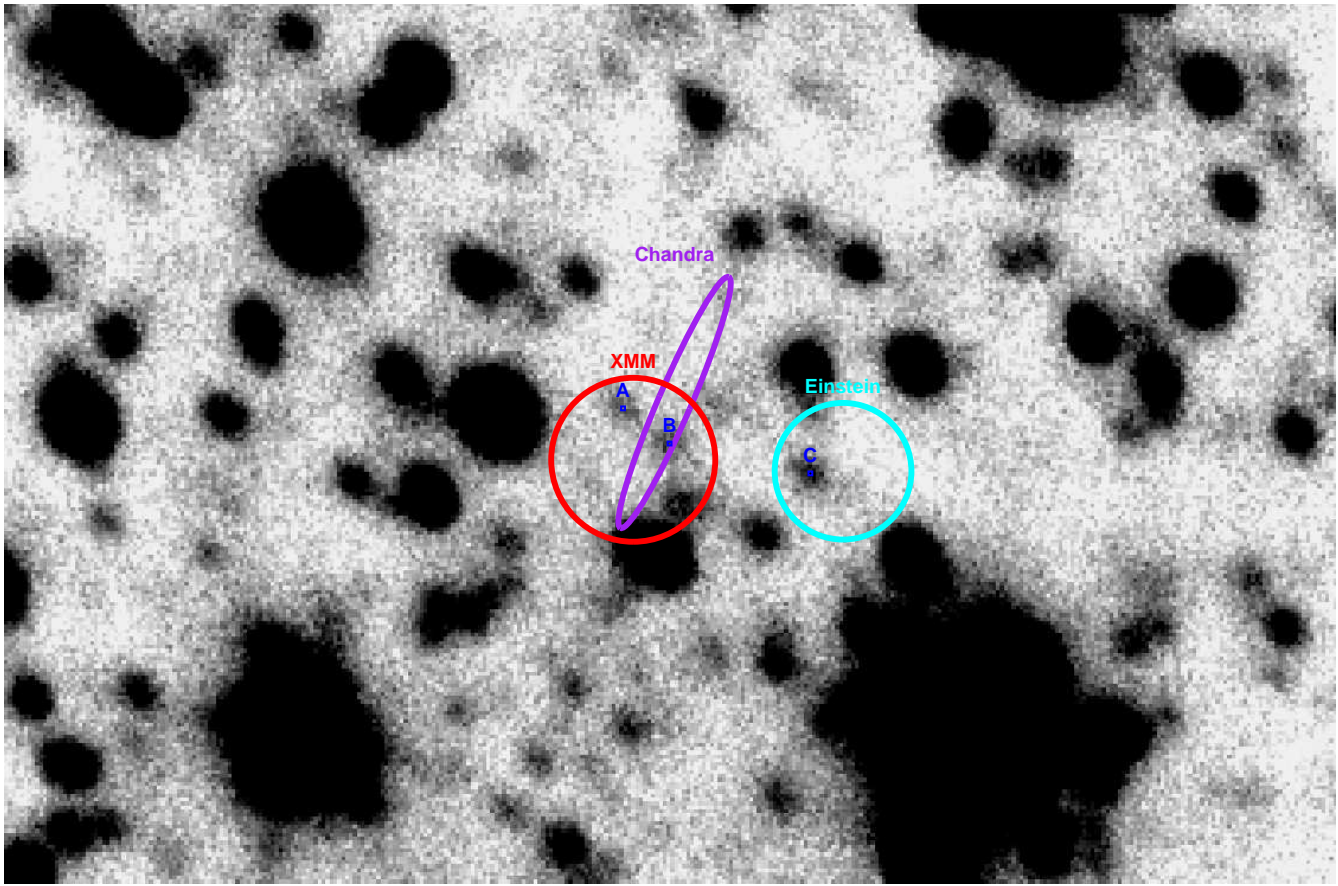}
 \caption{The co-added $K_s$ image of the 4U1323-619 field constructed
from 21 individual three second exposures after reduction and
correction for frame jittering. {\it Einstein}, {\it Chandra}, and
{\it XMM} positional uncertainties are overplotted. North is up,
East is left. Measured sources are designated by capital
letters; see the text for details.} 
 \label{error_boxes}
\end{figure*}

The infrared data analysed in this work were obtained on May 18, 2005
between 03:40 and 04:01 UT using the SOFI infrared  spectrometer and
imaging camera \citep{moorwood98} at the ESO 3.6m NTT telescope under
programme ID 075.D-0529(A). All scientific and calibration data were
retrieved through the publicly available ESO observational
archive\footnote{http://archive.eso.org}. The scientific data contain 21 three second exposures, each taken in Small Field imaging mode with pixel scale 0.144~arcsec~pix$^{-1}$ and seeing $\sim$1.1~arcsec.
Due to the specific nature of IR
observations, i.e. a rapidly changing background and
the techniques used to determine this, the reduction procedure is
quite different from optical imaging observations. To get the data
ready for scientific analysis we, therefore, used tools and recipes
provided by ESO: {\it GASGANO v2.3.0} for
data organisation tasks and the SOFI data reduction recipes from the {\it
Common Pipeline Library v5.0.0} to correct for bias, flat field and
frame jittering. After co-adding 21 reduced co-aligned frames (see Fig.~\ref{error_boxes}), 
we performed aperture photometry measurements with the {\it SExtractor} software
\citep{bertin96} and calibrated instrumental magnitudes using 2MASS
$K_s$ photometry of field stars. Our photometric errors therefore
include calibration dependency uncertainties of 0.10~mag. The astrometric
solution was obtained in the 2MASS reference frame with the {\it SCAMP}
software \citep{bertin06} and had 0.2~arcsec calibration uncertainty. 
Infrared measurements of all sources of interest in the 4U1323-619 field are listed in
Tab.~\ref{sources_params}.

\begin{table}
\caption{Positions and $K_s$ magnitudes of the sources from 
Fig.~\ref{error_boxes}. The astrometric uncertainty is 0.2~arcsec for all 
objects. Source B is the proposed new IR counterpart for 4U1323-619, marked 
boldface.}
\label{sources_params}
\begin{tabular}{lcccc}
\hline
Source & RA (J2000) & Dec (J2000) & $K_s$ \\
\hline
A & 13:26:37.22 & -62 08 07.7 & $18.56 \pm 0.23$ \\
{\bf B} & {\bf 13:26:36.98} & {\bf -62:08:09.0} & $\bmath{18.12 \pm 0.20}$ \\
C & 13:26:36.25 & -62:08:10.1 & $17.87 \pm 0.18$ \\
\hline
\end{tabular}

\end{table}

\subsection{Results}

We overplotted positional uncertainties obtained from the {\it XMM}
and {\it Chandra} data on the combined IR image (see
Fig.~\ref{error_boxes}), together with the {\it Einstein} error circle
which, following \citet{smale95}, is centred at the X-ray position of
Source D from \citet{parmar89}, as measured by {\it Einstein} HRI:
$\alpha_{J2000} = 13^h26^m36.08^s,~\delta_{J2000} =
-62\degr08\arcmin10.2\arcsec,~R = 2.5$~arcsec (90~per~cent
confidence). In Fig.~\ref{error_boxes} we designated sources of the
interest by a capital letters and measured their magnitudes and the
astrometric positions which are given in
Tab.~\ref{sources_params}. The single object within the {\it Einstein}
error circle (source C in this work) was proposed by \citet{smale95}
as the IR counterpart of 4U1323-619 and can now be ruled out. This mis-identification was probably caused by the underestimation of the position uncertainty radius for the source, which is quite faint for {\it Einstein}/HRI. We propose the only source visible inside {\it Chandra} and {\it XMM}
positional errors intersection down to a limiting magnitude $K_s \sim
19.4$~mag ($3\sigma$ upper limit), source B, as a new counterpart for
4U1323-619.

\section{Discussion}

We obtained simple estimations
of optical luminosity of the accretion disc illuminated by the central
isotropic spherical X-ray source with a 0.1-10~keV luminosity $L_X
\sim 5.2 \times 10^{36}$~erg~s$^{-1}$ \citep{boirin05} assuming for
convenience that its radius is $\sim {\rm few}\times10 R_g$ which
means that it is relatively small comparing to the accretion disc
size. While it is known that the source has varied in luminosity over
a 20 year period \citep{balucinska09} and an interpolation to the date
of the IR observations gives $\simeq 2.5$ times smaller luminosity, we leave
mentioned $L_X$ value because the bolometric luminosity is expected to
be the same factor higher. Given a system period $P=2.94$~h 
and assuming a compact object to be a neutron star with $M_X = 1.4M_\odot$,
one can estimate the mass of the secondary component from the
mass-radius relation for 
main sequence stars and the constraint that the star fills its Roche
lobe of size \citep{eggleton83}: 
\begin{equation}
\label{eq_radius}
 r/a = 0.49 \frac{q^{2/3}}{0.6 q^{2/3} + \ln(1 + q^{1/3})}
\end{equation}
where $q = M_{opt}/M_X$ and $a$ is the semi-major axis of the binary. This
simple estimate gives $M_{opt} \simeq 0.25M_\odot$ and enables us to
determine the accretion disc size using \citet{paczynski77}:
$r_{out} \simeq 0.51a = 0.62R_\odot$ for $\mu = M_{opt} / (M_{opt} +
M_X) = 0.15$.

The effective temperature of an accretion disc with a height
$z_0 \propto r^n$ illuminated by a point source in its centre is approximated by \citep{shakura73}: 
\begin{equation}
\label{eq_temp}
 \sigma T_{eff}^4 = \eta \frac{L_X}{4 \pi r^2} \cos \theta = \eta \frac{L_X}{4 \pi r^2} (n - 1) \frac{z_0}{r}
\end{equation}
where $\eta$ is the absorbed fraction of the radiation impinging on
the disc surface assumed to be $\sim 0.5$. Let us consider
the situation when the illumination does not change much the structure of
a standard $\alpha-$disc. We therefore adopt standard $\alpha$-disc
parameters, $n = 9/8, z_0/r_{out} \simeq 0.05$, giving a temperature $T_{eff,~out} = 10500$~K at the outer disc radius $r_{out}$. 

The observed flux (from a single side of the accretion disc as visible from Earth) 
at frequency $\nu = 1.4 \times 10^{14}$~Hz for the $K_s$ filter is close to 
the Rayleigh-Jeans range and can be estimated as follows: 
\begin{equation}
\label{eq_flux}
 Q_\nu = \frac{64}{49} \pi \cos i \frac{2 \nu^2 kT_{eff,~out} r^2_{out}}{c^2 d^2 f} e^{-\tau_\nu}
\end{equation}
Here the factor $64/49$ accounts for the $T_{eff}(r) \propto r^{-15/32}$
dependency in the standard accretion disc, and $f \simeq 1.3$ compensates
for a non Rayleigh-Jeans law for a given $h\nu/kT_{eff,~out} = 0.58$
ratio. Moreover, the observed flux is obviously decreased by the interstellar
extinction with optical depth $\tau_\nu$. Taking a moderate orbital
plane inclination angle $i = 70\degr$, a source distance $d = 10$~kpc,
and a typical extinction $A_{K_s} \simeq 2.0$~mag ($\tau_\nu
\simeq 1.84$) derived from the 3D galactic extinction map in this
direction \citep{marshall06}, gives $Q_\nu = 2.1 \times
10^{-29}$~erg~cm$^{-2}$~s$^{-1}$~Hz$^{-1}$ which is an order of magnitude
less than the observed flux $(3.1\dots4.5) \times
10^{-28}$~erg~cm$^{-2}$~s$^{-1}$~Hz$^{-1}$ translated from the measured magnitude $K_s=18.12 \pm 0.20$~mag and its uncertainty using zeropoints from \citet{cohen03}. 

We also estimated the observed flux in another limiting case for an
illuminated disc with constant temperature along the $z$ coordinate
(isothermal disc). Following equation (A5) from \citet{vrtilek90}
with the same set of the system parameters used above, one gets $T_{eff,~out} =
10000$~K. Use of $T_{eff}(r) \propto r^{-3/7}$ in integrating across the disc for isothermal
model gives a factor of $14/11$ instead of $64/49$ in Eq.~\ref{eq_flux} and the observed flux $Q_\nu$ becomes a factor of 0.9 smaller than in the first case, considering also slightly different $T_{eff,~out}$.

There are several possible explanations for this discrepancy between
the observed and predicted fluxes. The existence of a hot ($2 \times 10^6$~K)
atmosphere above the outer regions of a disc \citep{jimenez02} can increase the observed flux at
a given frequency $\nu = 1.4 \times 10^{14}$~Hz by a factor of
$1.3\dots1.7$ if we naively consider the effect of increased $z_0/r_{out} \simeq 0.2$ ratio on Eq.~\ref{eq_temp}--\ref{eq_flux}. In such a hot atmosphere, the scattering of
X-ray photons on free electrons with consequent penetration to
sub-photosphere layers and thermalization takes place. This
increases disc illumination and hence brightens the optical emission. 

Also we assumed in the beginning the compact nature of the central X-ray source while it is now known that there exists extended accretion disc corona with the radius $\simeq 30 000$~km for adopted $L_X$ \citep{church04,balucinska09}. This changes the illumination geometry of the outer regions of a disc, namely $\cos \theta \simeq z_0/r$ in Eq.~\ref{eq_temp}, and thus increases $T_{eff,~out}$ and $Q_\nu$ by a factor of $(n - 1)^{-1/4} \simeq 1.7$, correspondingly.

Placing the object to the distance of 4--5~kpc instead of 10~kpc
suggested by \citet{parmar89} on the basis of observed bursts being sub-Eddington
would account for all inconsistency of predicted
vs. observed fluxes because of significantly reduced extinction in
this direction and the fact that $Q_\nu$ scales with distance as
$d^{-3/2}$ for a fixed X-ray flux (which follows immediately from
Eq.~\ref{eq_temp}--\ref{eq_flux} since $Q_\nu \propto T_{eff,~out}/d^2
\propto L_X^{1/4}/d^2$ while itself $L_X \propto d^2$ for the
mentioned condition). We therefore might get $Q_\nu = 2.0 \times
10^{-28}$~erg~cm$^{-2}$~s$^{-1}$~Hz$^{-1}$ in this case.  

There is also a possibility of existence of a $\sim2-3$~mag fainter
object inside the combined {\it XMM} and {\it Chandra} error box which could be the actual
counterpart of 4U1323-619, but this cannot be ruled out on the basis
of existing archival data and requires dedicated high spatial
resolution X-ray observations and
deeper phase-resolved follow-up
IR observations, probably with adaptive optics in order to reduce field
contamination by the source B. Discovery of the 2.94~h variability period in the IR source would allow to finally identify the 4U1323-619 counterpart.

\section{Conclusions}

On the basis of examination of archival ESO NTT data within the area of overlap of {\it XMM} and {\it Chandra} error regions, we have identified a probable candidate for the IR counterpart of 4U1323-619. Its observed $K_s$ magnitude significantly
differs from that predicted by a simple analytical model of an accretion
disc illuminated by a hot central spherical corona with a parameter set
available in the literature. While putting the object to 4--5~kpc
instead of assumed 10~kpc would explain the discrepancy, we encourage
high spatial accuracy X-ray observations (e.g. using {\it Chandra}/HRC) and 
deeper phase-resolved follow-up observations inside the resulting error
box listed here.

\section*{Acknowledgments}
The authors thank Pavel Shtykovskiy for his help in {\it XMM} data analysis and our referee, Mike Church, for his prompt and clear report which helped to improve this letter.
This research is based on observations made with ESO 3.6m NTT
telescope at the La Silla under programme ID 075.D-0529(A) and has
made use of the VizieR catalogue access tool, CDS, Strasbourg,
France. IZ and NS were supported by the Russian
Foundation for Basic Research, grant 09-02-00032.

\bibliographystyle{mn2e}
\bibliography{4u1323}

\begin{thebibliography}{}

\bibitem[\protect\citeauthoryear{{Ba{\l}uci{\'n}ska-Church}, {Church},
  {Oosterbroek}, {Segreto}, {Morley} \& {Parmar}}{{Ba{\l}uci{\'n}ska-Church}
  et~al.}{1999}]{balucinska99}
{Ba{\l}uci{\'n}ska-Church} M.,  {Church} M.~J.,  {Oosterbroek} T.,  {Segreto}
  A.,  {Morley} R.,    {Parmar} A.~N.,  1999, \aap, 349, 495

\bibitem[\protect\citeauthoryear{{Ba{\l}uci{\'n}ska-Church}, {Dotani},
  {Hirotsu} \& {Church}}{{Ba{\l}uci{\'n}ska-Church}
  et~al.}{2009}]{balucinska09}
{Ba{\l}uci{\'n}ska-Church} M.,  {Dotani} T.,  {Hirotsu} T.,    {Church} M.~J.,
  2009, \aap, 500, 873

\bibitem[\protect\citeauthoryear{{Bertin}}{{Bertin}}{2006}]{bertin06}
{Bertin} E.,  2006, in {Gabriel} C.,  {Arviset} C.,  {Ponz} D.,   {Enrique} S.,
   eds, Astronomical Data Analysis Software and Systems XV Vol.~351 of
  Astronomical Society of the Pacific Conference Series, {Automatic Astrometric
  and Photometric Calibration with SCAMP}.
pp 112--+

\bibitem[\protect\citeauthoryear{{Bertin} \& {Arnouts}}{{Bertin} \&
  {Arnouts}}{1996}]{bertin96}
{Bertin} E.,  {Arnouts} S.,  1996, \aaps, 117, 393

\bibitem[\protect\citeauthoryear{{Boirin}, {M{\'e}ndez}, {D{\'{\i}}az Trigo},
  {Parmar} \& {Kaastra}}{{Boirin} et~al.}{2005}]{boirin05}
{Boirin} L.,  {M{\'e}ndez} M.,  {D{\'{\i}}az Trigo} M.,  {Parmar} A.~N.,
  {Kaastra} J.~S.,  2005, \aap, 436, 195

\bibitem[\protect\citeauthoryear{{Church} \&
  {Ba{\l}uci{\'n}ska-Church}}{{Church} \&
  {Ba{\l}uci{\'n}ska-Church}}{2004}]{church04}
{Church} M.~J.,  {Ba{\l}uci{\'n}ska-Church} M.,  2004, \mnras, 348, 955

\bibitem[\protect\citeauthoryear{{Church}, {Reed}, {Dotani},
  {Ba{\l}uci{\'n}ska-Church} \& {Smale}}{{Church} et~al.}{2005}]{church05}
{Church} M.~J.,  {Reed} D.,  {Dotani} T.,  {Ba{\l}uci{\'n}ska-Church} M.,
  {Smale} A.~P.,  2005, \mnras, 359, 1336

\bibitem[\protect\citeauthoryear{{Cohen}, {Wheaton} \& {Megeath}}{{Cohen}
  et~al.}{2003}]{cohen03}
{Cohen} M.,  {Wheaton} W.~A.,    {Megeath} S.~T.,  2003, \aj, 126, 1090

\bibitem[\protect\citeauthoryear{{Eggleton}}{{Eggleton}}{1983}]{eggleton83}
{Eggleton} P.~P.,  1983, \apj, 268, 368

\bibitem[\protect\citeauthoryear{{Forman}, {Jones}, {Cominsky}, {Julien},
  {Murray}, {Peters}, {Tananbaum} \& {Giacconi}}{{Forman}
  et~al.}{1978}]{forman78}
{Forman} W.,  {Jones} C.,  {Cominsky} L.,  {Julien} P.,  {Murray} S.,  {Peters}
  G.,  {Tananbaum} H.,    {Giacconi} R.,  1978, \apjs, 38, 357

\bibitem[\protect\citeauthoryear{{Frank}, {King} \& {Lasota}}{{Frank}
  et~al.}{1987}]{frank87}
{Frank} J.,  {King} A.~R.,    {Lasota} J.-P.,  1987, \aap, 178, 137

\bibitem[\protect\citeauthoryear{{Jimenez-Garate}, {Raymond} \&
  {Liedahl}}{{Jimenez-Garate} et~al.}{2002}]{jimenez02}
{Jimenez-Garate} M.~A.,  {Raymond} J.~C.,    {Liedahl} D.~A.,  2002, \apj, 581,
  1297

\bibitem[\protect\citeauthoryear{{Marshall}, {Robin}, {Reyl{\'e}}, {Schultheis}
  \& {Picaud}}{{Marshall} et~al.}{2006}]{marshall06}
{Marshall} D.~J.,  {Robin} A.~C.,  {Reyl{\'e}} C.,  {Schultheis} M.,
  {Picaud} S.,  2006, \aap, 453, 635

\bibitem[\protect\citeauthoryear{{Moorwood}, {Cuby} \& {Lidman}}{{Moorwood}
  et~al.}{1998}]{moorwood98}
{Moorwood} A.,  {Cuby} J.-G.,    {Lidman} C.,  1998, The Messenger, 91, 9

\bibitem[\protect\citeauthoryear{{Paczynski}}{{Paczynski}}{1977}]{paczynski77}
{Paczynski} B.,  1977, \apj, 216, 822

\bibitem[\protect\citeauthoryear{{Parmar}, {Gottwald}, {van der Klis} \& {van
  Paradijs}}{{Parmar} et~al.}{1989}]{parmar89}
{Parmar} A.~N.,  {Gottwald} M.,  {van der Klis} M.,    {van Paradijs} J.,
  1989, \apj, 338, 1024

\bibitem[\protect\citeauthoryear{{Shakura} \& {Sunyaev}}{{Shakura} \&
  {Sunyaev}}{1973}]{shakura73}
{Shakura} N.~I.,  {Sunyaev} R.~A.,  1973, \aap, 24, 337

\bibitem[\protect\citeauthoryear{{Smale}}{{Smale}}{1995}]{smale95}
{Smale} A.~P.,  1995, \aj, 110, 1292

\bibitem[\protect\citeauthoryear{{Vrtilek}, {Raymond}, {Garcia}, {Verbunt},
  {Hasinger} \& {Kurster}}{{Vrtilek} et~al.}{1990}]{vrtilek90}
{Vrtilek} S.~D.,  {Raymond} J.~C.,  {Garcia} M.~R.,  {Verbunt} F.,  {Hasinger}
  G.,    {Kurster} M.,  1990, \aap, 235, 162

\bibitem[\protect\citeauthoryear{{Warwick}, {Marshall}, {Fraser}, {Watson},
  {Lawrence}, {Page}, {Pounds}, {Ricketts}, {Sims} \& {Smith}}{{Warwick}
  et~al.}{1981}]{warwick81}
{Warwick} R.~S.,  {Marshall} N.,  {Fraser} G.~W.,  {Watson} M.~G.,  {Lawrence}
  A.,  {Page} C.~G.,  {Pounds} K.~A.,  {Ricketts} M.~J.,  {Sims} M.~R.,
  {Smith} A.,  1981, \mnras, 197, 865

\bibitem[\protect\citeauthoryear{{White} \& {Swank}}{{White} \&
  {Swank}}{1982}]{white82}
{White} N.~E.,  {Swank} J.~H.,  1982, \apjl, 253, L61

\end{thebibliography}

\label{lastpage}

\end{document}